\documentclass[sigconf,authorversion,nonacm]{acmart}

\usepackage{multirow}
\usepackage{enumitem}
\usepackage[english]{babel}

\AtBeginDocument{%
  }

\begin{document}

\title{``This could save us months of work'' - Use Cases of AI and Automation Support in Investigative Journalism}

\renewcommand{\shortauthors}{Anonymous Authors}

\author{Besjon Cifliku}
\email{besjon.cifliku@cais-research.de}
\orcid{0009-0007-5081-9531}
\affiliation{%
  \institution{Center For Advanced Internet Studies}
  \city{Bochum}
  \country{Germany}
}

\author{Hendrik Heuer}
\orcid{0000-0003-1919-9016}
\email{hendrik.heuer@cais-research.de}
\affiliation{%
  \institution{Center For Advanced Internet Studies}
  \city{Bochum}
  \country{Germany}
}
\affiliation{%
  \institution{Bergische Universität Wuppertal}
  \city{Wuppertal}
  \country{Germany}
}
\renewcommand{\shortauthors}{Cifliku and Heuer}

\begin{abstract}
    As the capabilities of Large Language Models (LLMs) expand, more researchers are studying their adoption in newsrooms. However, much of the research focus remains broad and does not address the specific technical needs of investigative journalists. Therefore, this paper presents several applied use cases where automation and AI intersect with investigative journalism. We conducted a within-subjects user study with eight investigative journalists. In interviews, we elicited practical use cases using a speculative design approach by having journalists react to a prototype of a system that combines LLMs and Programming-by-Demonstration (PbD) to simplify data collection on numerous websites. Based on user reports, we classified the journalistic processes into data collecting and reporting. Participants indicated they utilize automation to handle repetitive tasks like content monitoring, web scraping, summarization, and preliminary data exploration. Following these insights, we provide guidelines on how investigative journalism can benefit from AI and automation.
\end{abstract}

\ccsdesc[500]{Human-centered computing~Empirical studies in collaborative and social computing}
\ccsdesc[500]{Human-centered computing~Empirical studies in HCI}

\keywords{Computational Journalism,Automated Journalism,Automation,AI,Large Language Models (LLMs),Programming-by-Demonstration}

\maketitle

\section{INTRODUCTION}

Investigative journalism uncovers truths and shapes public opinion \cite{Stray14092019, hamilton-democracy-2016, nugent2022investigative} by collecting, analyzing, and fact-checking vast amounts of data. Nevertheless, the exponential growth in available data poses significant challenges in promoting transparency and justice in the digital age. Broader technical expertise and significant efforts are required to process and acquire online information. Automation assists investigative reporting by analyzing large datasets, finding correlations, and recognizing hidden patterns in documents that might be overlooked otherwise \cite{Diakopoulos_book_2019}. Accordingly, this allows journalists to automate repetitive tasks, focus on framing better insights, and create newsworthy stories. 

\citet{diakopoulus2024_GenAI_Journalism_Ethics} report the widespread use of generative AI for content creation, editing, and translation, as well as for researching, coding, and sensemaking.  In addition, integrating AI agents with web surfing opens new doors to access digital resources. According to \citet{cools2022, Cools26082024}, automation is already incorporated in news collection, production, verification, and news-sharing processes. 

This study examines how investigative journalists integrate automation and AI into their daily workflows, focusing on automation tasks, practical use cases, and the limitations of the tools they already employ. In this work, we do not constrain AI to generative models, but we consider AI as any system capable of doing autonomous work. Our approach aims to identify end-user requirements and features to guide the design and development of an automation framework for investigative journalism. To the best of our knowledge, this paper is among the first to involve investigative journalists deeply in a tool-based co-creation process through user studies interviews.

We used an elicitation \cite{wikipedia_requirements_elicitation} prototype during the interviews to introduce journalists to the PbD paradigm using LLMs. Our study employs a speculative discursive design methodology \cite{Auger2013SpeculativeDesign, dunne2013speculative, kolovson24SpeculativeDesign} combined with user elicitation \cite{vatavu2016ElicitationUserStudies}, encouraging interviewees to reflect on the potential benefits when applying a similar artifact in their daily context. The elicitation prototype facilitates user interactions on the web utilizing LLMs by allowing users to intuitively demonstrate how they want to collect web data. The system then replicates these actions and suggests methods for retrieving similar hierarchical data. The goal is to empower users, regardless of background, to automate web-based tasks easily. We used a video of the prototype during interviews to illustrate web automation use cases, to showcase the prototype's features, and to elicit responses. We explored how this novel interaction with generative AI could be integrated into journalists' workflows and evaluated their need for such an automation system. We further asked participants to suggest practical and relevant software features applicable to their work. 

We applied the Human-in-the-Loop (HITL) paradigm, similar to \cite{Missaoui-journalists-expertise-dminr-2019}, involving investigative journalists as end-users in the design process to tailor the technical solutions to their needs. This approach guided our investigation of the following research question.

\begin{itemize}
    \item \textbf{RQ: In what use cases is automation or generative AI used in investigative journalism workflow?}
\end{itemize}

To answer this research question, we conducted eight qualitative  within-subject semi-structured interviews with investigative journalists in Germany. 

Briefly, our work contributes to (1) a qualitative analysis of investigative journalists perspectives on AI and automation, focusing on a novel approach to automating repetitive tasks through a combination of PbD and generative LLMs and (2) a taxonomy of potential automation use cases and solutions that researchers can use as a foundation for developing new tools. Lastly, we briefly discuss some insights on challenges when using automation in journalistic settings.

\section{BACKGROUND \& RELATED WORK}

Computational journalism combines data science, algorithms, and social science to support journalists in shaping stories when working with the continuously growing volume of data \cite{hamilton-turner-2009, cohen-turner-2011}.  \citet{Diakopoulos_book_2019} identified various newsroom applications of computational algorithms, such as social media monitoring, fact-checking, document analysis, and automated reporting. Previous studies have agreed on four key processes in news automation: verification, production, gathering, and distribution \cite{cools2022, Cools26082024}. In this study, investigative journalism's \cite{Wuergler26102023, van2005investigative} scope mainly focuses on information gathering and analyzing and not much on reporting.

Scholars like \citet{Stray14092019}  and \citet{Fridman_2023} explored the hurdles of integrating AI in newsrooms. They identified data access and availability as significant obstacles in investigative research due to inconsistency, lack of open-source data, and transparency issues. \citet{Stray14092019} described automated data preparation as a prospective future application. It is important to note that this study predates the release of OpenAI’s ChatGPT \cite{openai_chatgpt}.

Web automation has long been a valuable tool for data collection \cite{barman2016_ringer}.  As the web evolves and transforms, the automation ecosystem adapts in response. However, this introduces new layers of complexities as the data retrieval process requires more technical expertise \cite{Chasins2019}.  Practitioners must understand how to navigate increasingly tangled web structures to retrieve data \cite{Chasins2019}. Automation reduces technical knowledge needs, enabling users to collect data without a deep understanding of web architectures \cite{barman2016_ringer, Chasins2019}. PbD enables users to automate tasks through demonstrations, eliminating the need for manual coding \cite{cypher93-programming-demonstration}. In another direction, a large body of researchers is studying how LLMs allow users to query data or automate web interaction using natural language \cite{tang2024steward, lai2023_AutoWebGLM, gur2024realworldwebagentplanninglong} with a few open-source initiatives \cite{skyvern_website, LaVague, fabric}.

Emerging computational methods are increasingly supporting journalistic processes. For example, \citet{osti_10206645} support political reporters in identifying newsworthy voter locations during elections using data mining algorithms. \citet{Broussard02112015} developed an AI system to analyze data and identify opportunities or newsworthy investigative ideas related to public affairs, while \citet{park2016_commentIQ} automated news articles' comments moderation and provided analytics for better storytelling. Similarly, \citet{petridis23AngleKindling} worked on co-designing and prototyping a tool to explore different reporting angles using LLMs. \citet{Jamil2024OnlineDI-SocialLens} enabled journalists to query data and extract insights from multiple online sources.

\citet{Kasica_Dirty_Data_2023} thoroughly interviewed $36$ data journalists to explore how they work and process dirty data (i.e., incomplete, inaccurate, or inconsistent information \cite{dirtydata2025}). Their study provides a taxonomy of data processing from data collection to data dissemination in the context of data journalism \cite{Kasica_Dirty_Data_2023}. We follow a similar approach but prioritize revealing practical use cases rather than comprehensively analyzing the data workflow chain.

Rather than simply exploring automation trends in investigative journalism, we aim to create a framework to develop AI-powered tools specifically for journalists. This paper explicitly explores the journalists' perception of automation using PbD and LLMs. \citet{Fridman_2023} argue that a practice-focused approach, as proposed by \citet{Barroca2018}, enables a deeper understanding of how technology shapes the investigative newsroom environment.

\citet{Fridman_2023} bridged the gap between researchers and industry practitioners by collaborating with journalists, data scientists, and IT experts in a transdisciplinary workshop. Journalists defined research topics, scoped projects, and planned the work. Our study shares this collaborative approach with a stronger user-centered focus by deeply involving journalists in the problem formulation. Similarly, \citet{Missaoui-journalists-expertise-dminr-2019} highlighted the need for more interactive collaboration with journalists and identified opportunities and challenges while co-creating human-centered AI tools with journalists. They co-designed the "DMINR AI" \cite{macfarlane2022dminrtoolsupportjournalists} to automate claim verification and exploratory analysis.

\section{METHOD}

\begin{figure*}[ht]
    \centering
    \includegraphics[width=\textwidth]{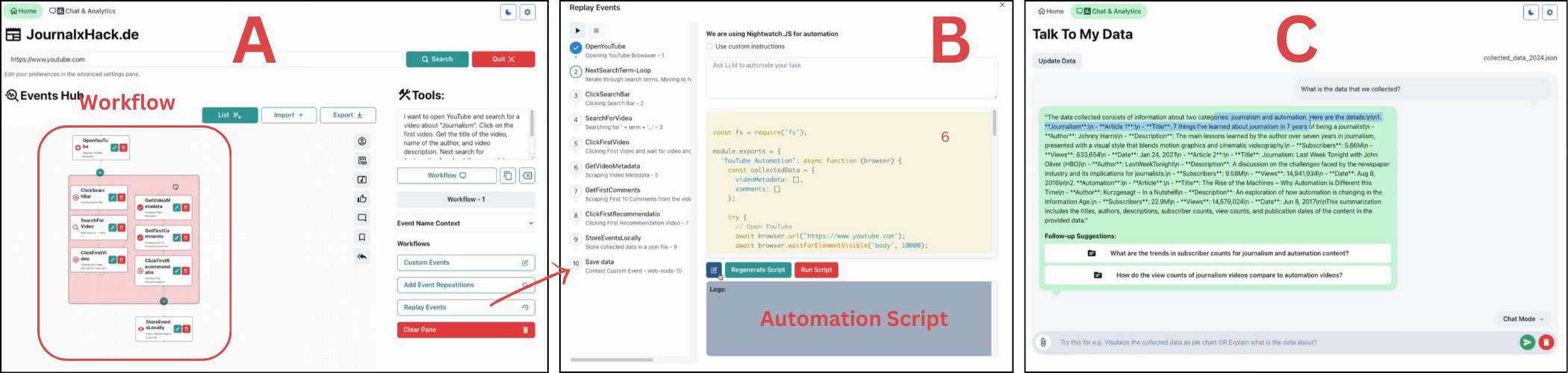}
        \label{fig:example_prototype}
    \caption{JournalXRecorder, prototype user interface. A) Main page where user can see their workflow, track events, and ask LLMs to automate. B) Running automation and generating scripts. C) Chat with and analyze the collected data. }
    \Description[JournalXRecorder UI]{JournalXRecorder, prototype user interface. A) Main page where user can see their workflow, track events, and ask LLMs to automate. B) Running automation and generating scripts. C) Chat and analyze the collected data.}
\end{figure*}

This paper aims to place journalists at the core of the research process, actively involving them in identifying areas where we can integrate automation into their tasks and engaging journalists with an elicitation prototype. The following section presents an overview of the participants, the prototype, and the procedure. 

\subsection{Participants}

We conducted within-subject user studies using semi-structured interviews. To recruit participants, we applied purposive \cite{Robinson2014} and snowball \cite{frey2018sage} sampling similar to \cite{Kasica_Dirty_Data_2023}. For our sampling strategy, we decided to recruit journalists with significant experience, particularly those with expertise in working with data. We contacted journalists through one of Germany's largest investigative journalism research networks, \href{https://netzwerkrecherche.org/ueber-uns/kurzportrait/}{\textit{Netzwerk Recherche}}. We initially recruited four participants from this channel. After each interview, we asked participants to recommend colleagues interested in joining the study. We contacted these referrals via email and continued reaching out until all recommendations were exhausted. Additionally, we recruited one more through university newsletters. The final sample comprised eight participants from major newspapers of record and news agencies across Germany. We did not provide monetary compensation for the participation.

The participants are listed in Table \ref{tab:participants}. The sample included five male and three female participants: three OSINT specialists, three data journalists, one state-funded cyber intelligence specialist, and one editor-investigative reporter.
Participants ranged in age from $34$ to $40$, with a mean of $37.5$ (SD$\mathcal{\approx}2.26$). Educational backgrounds included two doctorates, five master’s degrees, and one college degree. All participants had over five years of experience and were familiar with generative AI and automation. All participants, except for $\mathbf{P07}$, had programming background. 

\subsection{Elicitation Prototype: JournalXRecorder}\label{anchor:elicitatoin_prototype}

We designed an elicitation prototype around three main objectives. First, it should feature a \textbf{record and replay} functionality \cite{Chasins2019}, enabling journalists to capture and replay the web activity chronologically. Second, users must be able to automate and customize their workflows using LLMs \textbf{without requiring technical expertise}. Finally, users should be able to extract \textbf{structured data} from any website. For instance, journalists could use the tool to monitor how content on various sites evolves. They would only need to visit a site once, select the elements they wish to track, and the tool would then perform the checks repeatedly. Like \citet{Chasins2019}, the motivation is to \textbf{do it once and repeat it anytime, anywhere}. The tool lets users collect data online by recording their browser interactions, such as clicking, typing, and scrolling. To engage journalists with this automation framework, we employed speculative design \cite{dunne2013speculative, Auger2013SpeculativeDesign} using a video of the prototype as our key method.
 
\subsection{Procedure}

We instructed participants to prepare a use case that AI could automate. We began the interviews by obtaining participants' informed consent to use their data in compliance with GDPR and requesting permission to record the sessions. 

We divided the semi-structured interviews into two phases. In the first part (approximately 20 minutes), we asked participants about their familiarity with automation and AI. They responded to questionnaires regarding the impact of automation on their work, data management practices, and the challenges they face. Further, we questioned participants about the limitations of their current automation tools and their desired improvements to their workflows. Finally, we tasked participants to \textbf{imagine} an ideal tool tailored to their needs, considering both computational and human resources available for its creation. In a storytelling setting, we aimed to explore practical journalistic use cases and point out any potential limitations.

In the second part, we presented the elicitation prototype. We listed its main features, explained a few details for the use case as illustrated in Appendix \ref{anchor:example_use_case}, and presented an eight-minute demo in the form of a video. Afterward, we requested participants to suggest potential scenarios for its application during investigations.

Interviews lasted, on average, $62$:$39$ minutes (SD$\mathcal{\approx}9$:$10$) and were conducted virtually via Webex. To transcribe the audio, we used an automatic caption generator provided by Webex. In addition, the first author noted down key points during the interviews. We applied qualitative content analysis to identify key themes through inductive coding following \cite{Mayring1991, Mayring2000}. We reviewed the interviews iteratively, grouping codes into a set of overarching themes similar to \cite{heuer-glassman-2023-accessible-tools}. The first author assigned the preliminary codes. Following the interviews, we identified potential concepts related to news processes. We then mapped these categories to the participants' statements and engaged in an iterative process, repeatedly revisiting the participant's statements to draw new codes \cite{corbin2014basics} related to news automation. The codes were subsequently merged and refined into themes and sub-themes. The first author did the coding, while the second author reviewed the codes in weekly sessions to ensure the themes' relevance to the participants' statements. The interview questionnaires can be found in Appendix \ref{anchor:example_use_case}

\section{RESULTS}

\renewcommand{\arraystretch}{1.5}

\begin{table*}[ht]
    \centering
    \small
    \Description[This table depicts the investigative journalism processes grouped into two categories: collecting and reporting data.]{This table depicts the investigative journalism processes grouped into two categories: collecting and reporting data.}
    \caption{This table depicts the investigative journalism processes grouped into two categories: collecting and reporting data. }
    \scalebox{0.9}{
    \begin{tabular}{|p{1.5cm}|p{2cm}|p{11cm}|}
        \hline
        \textbf{Process} & \textbf{Theme} & \textbf{Description} \\
        \hline
        \hline
        \multirow{3}{2cm}{\textbf{Collecting}}
        & Monitoring  &  Auditing, gathering, and tracking content information from multiple sources.
        \\
        \cline{2-3}
        & Filtering  & 
        Reducing redundancy and only targeting relevant information.
        \\
        \cline{2-3}
        & Documenting & 
        Recording of the investigation process and keeping track of each action.
        \\
        \cline{2-3}
        & Storing & 
        Preserving the collected data into usable format, converting it into searchable entities, and indexing into databases.
        \\
        \cline{2-3}
        & Augmenting & 
        Assisting in scaling up scraping tasks, providing analytical support and investigation leads.
        \\
        \hline
        
        \multirow{3}{2cm}{\textbf{Reporting}}
        & Analysing & 
        Simplifying and comprehending the results. Structuring and processing of data to draw insights.
        \\
        \cline{2-3}
        & Labeling  & 
        Categorizing, clustering, and linking data into themes and subtopics.
        \\
        \cline{2-3}
        & Writing & 
        Inspiration or assistance starting or clarifying tasks. AI news production and dissemination. \\
        \hline
\end{tabular}}
\label{tab:thematic_analysis}
\end{table*}

The following presents a taxonomy of automation scenarios and investigative journalism tasks based on systematically reviewed use cases. Interviews revealed that automation facilitates news  \textbf{collection} and \textbf{reporting}, as listed in Table \ref{tab:thematic_analysis}. 

\subsection{Collecting}

\subsubsection{Monitoring}

\textbf{Continuous Analytics and Alerting:} Participants emphasized using automation for continuous data collection and generating comprehensible responses, often through web scraping ($P01, P02, P03, P06, P07, P08$). For instance, $P02$ analyzed hotel prices during the European soccer championship and tracked climate change trends by comparing historical temperature records across cities. Moreover, $P03$ and $P08$ reported on downloading and summarizing data from Telegram groups. In addition, $P08$ recounted collecting metadata from GitHub to verify content and visualize repository activities. Furthermore, $P06$ suggested using the prototype for notifying the public about topic monitoring on local democracy. $P08$ proposed an automatic analysis of how  influencers on YouTube shift their language over time to study radicalization patterns. Additionally, $P06$ described how combining and reporting data from local air sensors can alert the public about air pollution.

\textbf{Tracking:} 
Participants also reported on tracing leaked information and tracking suspicious user fingerprints across online platforms, social media, public databases, and geospatial locations ($P01, P04, P05, P08$). For instance, $P01$ described tracking ``personas of interest'' on foreign-language websites or leaked documents, noting the process complexity and the lack of a centralized data storage. He suggested that an AI assistant capable of understanding initial hints and automating research could \textit{``save months of work''} and \textit{``make more interesting stories possible.''} Similarly, $P05$ recognized the potential of automation to process and extract key details from large document sets, providing an example of comparing websites' hidden Google tags to trace their origins. Likewise, $P08$ discussed comparing and tracking IP addresses to verify if sites are related. In another case, $P04$ pondered on having precise tools that provide greater control when geolocating images. She clarified that \textit{``If you got a picture in Gaza and if you try to find one house, it takes hours''}.

\textbf{Content Audit Trail:} Some participants noted that automation could be used to compare and track historical changes in online content to verify if it was maliciously altered ($P05, P08$). $P05$ described downloading complete website archives using Wayback Machine \cite{internet_archive} and matching their content against search phrases. Similarly, $P08$ highlighted the potential of using the prototype to monitor social media content for periodic changes. Furthermore, he expressed the idea of \textit{``taking snapshots of''} news websites and analyzing their content every hour to check how different outlets report particular news events.

\textbf{Universal Data Scrapers:} Participants envisioned a universal scraper capable of retrieving data using natural language ($P01, P06,$ $P07, P08$), particularly useful when journalist coverage is limited ($P07$). For instance, $P06$ and $P07$ highlighted the challenge of retrieving unstructured, non-standardized regulations from local state governments across Germany, where documents are inconsistently formatted and lack systematic accessibility. $P06$ emphasized the importance of city council proceedings for local democracy, noting the absence of a coherent publishing system. Similarly, $P07$ remarked, \textit{``I have to check around 18 different websites regularly, and it would be great to have a tool to collect that data''}. Additionally, $P07$ suggested using such a tool to monitor the social media accounts of radical political party members. In addition, $P01$ noted its potential for retrieving data from opaque social media platforms, which he referred to as \textit{``black boxes''}. Both $P01$ and $P08$ endorsed the idea of multi-modal scrapers capable of transcribing, analyzing, and summarizing speech, audio, or video content.

\textbf{LLMs Web Search Engines:} Participants ($P01, P03$) suggested AI agents functioning as advanced search engines. $P03$ noted that posing a question to a \textit{chatbot} is far easier than manually searching for answers online. Similarly, $P01$ explained that much of their work involves \textit{``googling and finding relevant details''} and argued that generative AI agents simplify the retrieving process and utilizing information.

\subsubsection{Filtering} Participants also recognized the value of LLMs in reducing redundant information $(P01, P02, P03$, $P05, P07)$. For example, $P03$ shared how she used ChatGPT to clean web markups, spotlight elements, and generate automation scripts. Furthermore, $P02$ and $P07$ highlighted its application in searching documents and automated findings reporting. Additionally, $P01$ primarily relied on automation for fact-checking and document analysis. However, automated fact-checking using machine learning is still limited, as agents struggle to distinguish between complex and trivial facts. $P04$ explained how automation could help when investigating economic topics such as white-collar crime (i.e., corruption) by looking into big data registries and quickly filtering and categorizing information. 

\subsubsection{Documenting} Participants also discussed the importance of documenting and maintaining a clear record of the investigative process. $P05$ and $P08$ reported that they usually do not work with much data but rather with small snippets of leaked information. Their job involves researching multiple websites to gather information and integrating it with government databases. As noted by $P05$, documenting this process could be challenging and time-consuming, and recalling the origin of the data may be difficult. $P05$ described a tool, which he referred to as \textit{``a time machine''}, that records users' digital activities during investigations and allows them to retrace their steps back to the source.

\subsubsection{Storing} During the interviews, participants highlighted challenges associated with data storage and cloud architecture ($P03$, $P05$, $P07$, and $P08$). $P03$ and $P07$ reported issues related to saving data in databases, which they attributed to a lack of software knowledge. $P05$ claimed that journalists often waste significant time copying data from multiple files. He reflected, \textit{``it takes your time to do great research''}, and speculated that an automated solution could solve this issue. Additionally, he suggested implementing an automated system to convert investigation reports into Excel files. Furthermore, $P08$  proposed an automated process for saving data to the cloud using LLMs.  Interestingly, $P03$ proposed sharing automation workflows among outlets without exchanging the collected data to prevent sensitive data leakage while calling a plugin that could automatically collect \textbf{any (un)structured} online content and store it in the cloud as  \textit{``a superpower''}.

\subsubsection{Augmenting} Participants believed that AI and automation assist in augmenting information discovery and boosting their productivity. $P08$ illustrated an automated process by expanding the scraping scripts, automatically executing in the cloud, storing collected data, and reporting the progress to the user. Interestingly, $P07$ considered the automatic email folder organization as an automation tool. All participants used generative AI to program and write scripts. $P04$ stated using LLMs to get initial investigation leads by analyzing small snippets during the investigation.

\subsection{Reporting}

\subsubsection{Analyzing} Themes on the role of data exploration using LLMs emerged repeatedly during interviews ($P01, P02,$ $P07, P08)$. Participants $P02, P03$, and $P08$ warned about hallucinations \cite{huang_hallucination_2023} when using LLMs. Despite these concerns, their ability to quickly identify patterns overlooked during research makes them useful data visualizers ($P02$, $P07$, and $P08$). For example, $P03, P04$, and $P05$ underlined the idea of representing data in knowledge graphs. $P05$ stated that an AI-assisted KG could potentially optimize the query results.
Furthermore, $P01$ elaborated on generating info-graphics metrics. In addition, $P01$ briefly explained a concept of a large RAG \cite{lewis_rag} using several LLMs to securely chat with $150$ million various formatted articles from their archive. $P04$ mentioned using LLMs to generate initial leads when verifying leaks before beginning the investigation. $P03$ also argued using LLMs to generate relevant queries for further investigation. In contrast, $P03$ was doubtful about the benefit of LLMs in data analysis and raised concerns about deterring data literacy. Analogously, $P06$ stated, \textit{``I am just not sure if it [LLMs] can interpret data in the way that humans do''}.

\subsubsection{Labeling} Participants also mentioned clustering content and discovering relations ($P01, P06$). $P01$ was already working on automatic data annotating and labeling articles using generative AI technologies. Interestingly, $P06$ explains the process of creating a crime map network based on police reports. 

\subsubsection{Writing} Participants also used LLMs to summarize text from documents or transcribed interviews. $P06$ reported using LLMs to rewrite hundreds of parents' reports on dangerous children's school routes and generate news headlines.  $P07$ considered using LLMs for \textit{``brainstorming inspiring ideas''} while writing reports but refrained from using them for rephrasing or copy editing, as he valued the originality of his work. Additionally, $P06$ mentioned that LLMs are helpful for text editing, such as citing interview question outlines, and noted that automation can also be applied to storytelling. He touched on using an AI illustrator and discussed integrating a print layout automation system. 

\section{DISCUSSION}

The following section outlines the study outcomes, discussing how automation and AI impact digital investigation. We find that investigative journalists must work with vast data sources while ensuring the reported information is still valid, accurate, and reliable. They often use automation as a means to build custom workflows for data extraction from public databases, social media platforms, or proprietary document archives.

We provide a set of journalistic use cases that can act as a basis for developing new automation tools. Our findings showed that the reported use cases are mainly related to information gathering rather than reporting. These results may be influenced by participants' OSINT-based profiling and the focus of the elicitation prototype on data scraping and analysis. Further studies on understanding reporting scenarios could help to complete the entire automation pipeline. 

\subsection{Automation Helps When Done Right}

With the growing scope of automation and adoption of LLMs in journalism, more tangled stories can be investigated. Our interviews indicate that many tools are already available to journalists, and new ones will continue to emerge. However, we find that many existing research solutions do not adequately address user needs. Therefore, we emphasize the importance of involving experts, affected stakeholders, and developers in a co-creation process to enhance the transparency of such systems. 

Our study reveals that a demonstration-based LLM system could simplify information extraction by automating repetitive tasks, significantly speeding up research.  Participants noted the need for multi-modal agents to facilitate source monitoring. To this end, stacking agents with different capabilities could reduce the workload, broaden automation scope, and accelerate investigation outcomes, especially when dealing with unstructured information. We reported that journalists without sufficient technical training might struggle with cloud storage. LLMs can streamline structuring data, converting files on multiple formats, and simplifying data ingestion. In addition, our findings underscore the importance of an automated systematic documentation process to maintain investigation records, as it allows journalists to trace back investigation leads, enabling more transparency. Furthermore, researchers can explore alternative methods to share automation workflows as described by $P03$, which can encourage collaboration among news outlets and enhance privacy. Another positive aspect of this approach is that it improves the reproducibility of online research, provided that the data is accessible.

To summarize, we can identify a few design recommendations based on study results for future research. Users should be able to (1) get (un) structured data quickly using multi-modal agents running periodically; (2) automatically convert and store collected data in multiple formats with the support of LLMs; (3) document, record, and track the actions during investigations; and (4) securely share automation flows without sharing sensitive data. 

\subsection{Ironies When Automating Investigations}
 
On another level, while automation boosts efficiency, it introduces new risks related to reliability, accountability, and transparency \cite{komatsu-ai-embody-values-2020}. Journalists value accurate and credible story reporting.

The \textbf{bias} embedded in the training data used to power the AI systems \cite{jarke2024blackbox} can diminish journalistic values \cite{komatsu-ai-embody-values-2020} as it can generate inaccurate stories that can mislead or radically influence the audiences. Additionally, \textbf{LLMs hallucinations} \cite{huang_hallucination_2023} can produce inaccurate facts, undermining the trustworthiness of such systems and the reliability of their output. This way, automation could harm the reputation of the publishers~\cite{graefe2016guideAutomatedJournalism} and also threaten consumers' trust. For example, when doing web automation, LLMs may generate code with outdated dependencies, introducing vulnerabilities and security issues $(P08)$.  Raising awareness, \textit{``AI literacy''}, and establishing clear guidelines for using AI in the newsroom are essential for building trust in autonomous systems. This  could be achieved by organizing workshops within investigative teams or developing simple prototypes to demonstrate and test system capabilities.
 
We identified several use cases affected by automation, in particular, related to information retrieval and source monitoring. Retrieving data from complex digital systems is complicated and requires knowledge and expertise. Consequently, the overlap between data science and investigative reporting continues to expand \cite{diakopulous2024_data_journalism}. Even if outlets automate fact-checking~\cite{wolfe2024ImpactOportunitiesGenAIFactchecking} through automatic claim verification, human intervention would still be necessary to validate and ensure the relevance and accuracy of results. While automation helps in information retrieval and could bridge knowledge gaps, journalism  involves more than just collecting data. The ability of journalists to structure and fact-check sources and evaluate automation outcomes utilizing computational and data thinking ~\cite{diakopulous2024_data_journalism} is crucial in preserving the integrity of automated reporting. Additionally, journalists need statistical knowledge and a deep understanding of how information flows on the internet to identify suitable data, interpret it, and create the right stories $(P03)$ . \citet{kouts-klemm2019} argues that the information loses value unless news workers properly interpret it. Due to the shift in today's data-driven digital world, there is an ongoing recalibration of responsibilities related to journalistic roles in the newsroom \cite{graefe2016guideAutomatedJournalism, diakopoulos2024generativeLabor}. Journalists must acquire new skills to uphold technology updates~\cite{diakopulous2024_data_journalism} and embrace the idea of \textbf{data and internet literacy}, which refers to the ability to structure, analyze, understand, and report insights from data ~\cite{friedrich2024dataliteracy}. 

Another topic is \textbf{budget constraints}, particularly faced by local news agencies \cite{Farhi2024NewsroomShrinkage, komatsu-ai-embody-values-2020}, when implementing large-scale automation $(P06)$. Running scrapers on the cloud or paying third-party services to host automation systems can be costly. A local outlet may lack the resources to support such tools for finding newsworthy stories $(P06)$. Furthermore, an investigation topic  might not be \textbf{relevant} for a long time to be automated, and specific topics might show up only once \cite{Stray14092019} $(P05)$.

\subsection{Limitations} The study's limitations include a small number of participants based in Germany, with an average age of $37.5$, as we focused on recruiting experienced practitioners. We leveraged the practitioners' technical backgrounds and expertise to inform our future goal of transforming their workflows into tools targeting less technically trained news workers. A larger participant pool, including more non-technically skilled journalists, could reveal additional automation use cases. In addition, we did not let users interact with the prototype, which could have yielded more insights.

\section{CONCLUSION}

This paper serves as a starting point for exploring how generative AI and PbD can assist journalists in data collection and automation processes. We provide actionable insights on \textbf{collection} and \textbf{reporting}. 

Our findings highlight the potential for journalists to adopt automation technologies that simplify data retrieval from multiple sources without the need for programming skills. We encourage future work to examine the potential of PbD in the context of automating journalistic workflows.

\bibliographystyle{ACM-Reference-Format}
\bibliography{references}

\appendix
\section{Appendix}

\subsection{Participants}

\begin{table}[h!]
\small
\caption{Distribution of participants according to their role at the newsroom, their age, gender, and their educational background}
\Description[Distribution of Participants]{Distribution of participants according to their role at the newsroom, their age, gender, and their educational background}
\scalebox{0.95}{
\begin{tabular}{|l|l|l|l|l|}
\hline
\multicolumn{1}{|c|}{} & \multicolumn{1}{c|}{\textbf{Role}} & \multicolumn{1}{c|}{\textbf{Age}} & \multicolumn{1}{c|}{\textbf{Gender}} & \multicolumn{1}{c|}{\textbf{Education}} \\ \hline
P1                               & OSINT (Fact Checking)              & 39                                & m                                    & PhD                                     \\ \hline
P2                               & Data Journalist                    & 34                                & f                                    & Master                                  \\ \hline
P3                               & OSINT (Fact-checking)              & 37                                & f                                    & Master                                  \\ \hline
P4                               & Data Journalist                    & 34                                & f                                    & Master                                  \\ \hline
P5                               & OSINT (Digital Forensic)           & 37                                & m                                    & Master                                  \\ \hline
P6                               & Data Journalist                    & 40                                & m                                    & PhD                                     \\ \hline
P7                               & Reporter \& Editor                 & 39                                & m                                    & Master                       \\ \hline
P8                               & Cyber-intelligence          & 40                                & m                                    & College Degree                          \\ \hline
\end{tabular}
}
\label{tab:participants}
\end{table}

\subsection{Automation Use-Case Video Prototyping}
Illustration of a use case scenario to audit data from YouTube similar to \citet{heuer_youtube_2021} using the elicitation prototype explained in Section \ref{anchor:elicitatoin_prototype} .\label{anchor:example_use_case}
\begin{quote}
   The user starts inputting a URL in the prototype, which triggers a local browser to pop up. The user navigates to YouTube and searches for ``Journalism''. Upon clicking the first search result, the user is redirected to the corresponding video page. The user can then select elements to scrape, such as the video title, description, or the number of comments. The user can also click on a second video recommendation. Throughout the process, JournalXRecorder tracks all interactions.
   While the user only needs to demonstrate the scraping process once, the tool can automatically collect all video results. Once the scraping is complete, the user uploads and analyze the collected using the chat.
\end{quote}

\subsection{Interview Questions}
\begin{enumerate}
    \item Can we record the video?
    \item We would like to know a bit more about you. Could you please introduce yourself, explain what you do, and why you are interested in this workshop? What news outlets have you worked for in the past?
    \item Could you share your gender, age, and highest education level?
    \item  What kind of journalism are you working on?
    \item For which tasks do you use Chat GPT or other similar AI based tools for?
    \item How do these AI tools affect your work?
    \item Have you ever collected and analyzed large amounts of data? If so, for which tasks?
    \begin{itemize}
        \item Did you find it challenging to do? 
        \item If yes, could you please elaborate on the challenges or limitations? 
    \end{itemize}
    \item Have you used any automation (tools) for your work?
    \begin{itemize}
        \item YES: What kind of automation do you use?
        \item YES: Why did you use automation?
        \item YES: How did automation affect your work?
        \item YES: What challenges did you face when automating?
        \item NO: Why did you not use any automation?
    \end{itemize}
    \item What challenges or limitations do you face when you want to use automation tools? 
    \item Imagine having all the necessary computational and human resources available; what kind of tool (with or without AI) do you wish for that could support you on your job? \\
\end{enumerate}

After the participants had seen the prototype, we further asked:
\begin{enumerate}
    \item What similar tools have you seen or used before?
    \item What are your thoughts on our prototype?
    \item What did you like about the tool? 
    \item What did you dislike about the tool? 
    \item How could this tool assist you in your work? 
    \item Could you think for a few minutes about use cases where you could use such a tool in your work? If not, why would you not use it?
    \item From your perspective, we would love to know what we can do to improve our tool. What new features would you like to have? 
    \item Open discussion
\end{enumerate}

\end{document}